\documentclass[a4paper,11pt]{article}

\pdfoutput=1 
\usepackage{jcappub}

\usepackage{graphicx}
\usepackage{longtable}
\usepackage{float}
\usepackage{dcolumn}
\usepackage{bm}
\usepackage{appendix}
\usepackage{multirow}
\usepackage{color}
\usepackage[utf8]{inputenc}
\usepackage{footmisc}

\newcommand{\mytilde}{\raise.17ex\hbox{$\scriptstyle\mathtt{\sim}$}}
\newcommand{\barr}{\begin{eqnarray}}
\newcommand{\earr}{\end{eqnarray}}
\newcommand{\bea}{\begin{eqnarray*}}
\newcommand{\eea}{\end{eqnarray*}}
\newcommand{\beq}{\begin{equation}}
\newcommand{\eeq}{\end{equation}}

\renewcommand{\bf}{\rm}
\title{Dark sector interactions and the curvature of the Universe in light of Planck's 2018 data}

\author[a,b]{M. Benetti}\emailAdd{micol.benettim@unina.it}

\author[c]{H. A. Borges}\emailAdd{humberto@ufba.br}

\author[c]{C. Pigozzo}\emailAdd{cpigozzo@ufba.br}

\author[c,d]{S. Carneiro}\emailAdd{saulo.carneiro.ufba@gmail.com}

\author[e]{J. S. Alcaniz}\emailAdd{alcaniz@on.br}

\affiliation[a]{Dipartimento di Fisica  ``E. Pancini", Universit\`a di Napoli  ``Federico II", Via Cinthia, I-80126 Napoli, Italy}

\affiliation[b]{Istituto Nazionale di Fisica Nucleare (INFN), sez. di Napoli, Via Cinthia 9, I-80126 Napoli, Italy}

\affiliation[c]{Instituto de F\'{\i}sica, Universidade Federal da Bahia, 40210-340 Salvador, BA, Brasil}

\affiliation[d]{PPGCosmo, CCE, Universidade Federal do Esp\'irito Santo, 29075-910 Vit\'oria, ES, Brasil}

\affiliation[e]{Departamento de Astronomia, Observat\'orio Nacional, 20921-400 Rio de Janeiro, RJ, Brasil}

\abstract{We investigate the observational viability of a class of interacting dark energy (iDE) models in the light of the latest Cosmic Microwave Background (CMB), type Ia supernovae (SNe) and SH0ES Hubble parameter measurements. Our analysis explores the assumption of a non-zero spatial curvature, the correlation between the interaction parameter $\alpha$ and the current expansion rate $H_0$, and updates the results reported in \cite{micol}. Initially, assuming a spatially flat universe, the results show that the best-fit of our joint analysis 
clearly favours a positive interaction, i.e., an energy flux from dark matter to dark energy, with $\alpha \approx 0.2$, while the non-interacting case, $\alpha = 0$, is ruled out by more than $3\sigma$ confidence level. On the other hand, considering a non-zero  spatial curvature, we find a slight preference for a negative value of the curvature parameter,  which seems to relax the correlation between the parameters $\alpha$ and $H_0$, as well as between $H_0$ and the normalization of the matter power spectrum on scales of 8$h^{-1}$ Mpc ($\sigma_8$). 
Finally, we discuss the influence of considering the SH$0$ES prior on $H_0$ in the joint analyses, and find that 
such a choice does not change considerably the standard cosmology predictions but has a significant influence on the results of the iDE model.} 

\begin{document}
\maketitle

\section{Introduction}
\label{Introduction} 
The increasing tension between the local measurements of the current  expansion rate \cite{friedman,Riess:2019cxk} and that derived from the temperature anisotropies in the Cosmic Microwave Background (CMB) \cite{Aghanim:2018eyx} assuming the $\Lambda$-Cold Dark Matter ($\Lambda$CDM) model has motivated the investigation of generalized models of the dark sector, physics beyond the standard model,  and alternative gravity theories \cite{Verde:2019ivm, Graef:2018fzu, Benetti:2017juy, DiValentino:2019qzk, Bernal:2016gxb, Guo:2018ans, Vattis:2019efj, Capozziello:2020nyq, Benetti:2020hxp, Benetti:2019gmo, Pan:2019gop,Poulin:2018cxd,Knox:2019rjx,Yang:2021hxg,Alcaniz:2019kah}. In the case of dynamical dark energy models, an important aspect worth considering is how to properly treat the dark energy (DE) perturbations,  
which in principle can be done in two different ways. 
The first one is to explicitly include  DE perturbations in the perturbed equations and assume a sound speed to DE. If a luminal sound speed is assumed DE perturbations can be always neglected. An alternative procedure is to decompose the dynamical dark energy into a pressureless, clustering component and a vacuum-type term with an Equation-of-State (EoS) parameter, $w = -1$ \cite{Zimdahl:2005ir, Wang:2013qy, Borges:2013bya}. In this case, it is possible to show \cite{micol} that the latter does not cluster in the limit of sub-horizon scales, while the former can be reinterpreted as dark matter. In this framework, a flux of energy between the dark components generally occurs, whose observational signature would indicate a dynamical nature of the original DE field. Such an energy flux violates adiabaticity and characterizes the so-called interacting DE models (iDE).

In a previous work \cite{micol}, it was shown that a joint analysis of the CMB (Planck 2015) data and SNe observations 
constrains the interaction parameter $\alpha$ of a particular class of iDE models to be slightly positive, corroborating results of similar studies \cite{wands2,Aurich:2017lck}. Moreover, a strong correlation was found between $\alpha$, the Hubble constant ($H_0$), and the normalization of the matter power spectrum on scales of 8$h^{-1}$ Mpc ($\sigma_8$), i.e., while a positive interaction parameter favours higher values of $H_0$, a negative $\alpha$ favours lower values of $\sigma_8$. This correlation is particularly important in the study of iDE models as a possible solution of the current $H_0$ and $\sigma_8$ tensions. 

The aim of this paper is twofold: first, to perform an updated analysis of \cite{micol} with the Planck (2018) likelihoods \cite{Aghanim:2018eyx}, where the CMB polarization has been taken into account. Second, to explore the influence of a non-zero spatial curvature in the determinations of the other model parameters, motivated by the recent study of \cite{DiValentino:2019qzk}. The present analysis also pays special attention to the role played by the $H_0$ prior from local measurements on the parameter estimates. In particular, we find that such a prior has a significant influence on the determination of the iDE model parameters, given the $\alpha$-$H_0$ correlation reported in \cite{micol}.

\section{Parametrising the interaction}
\label{Sec:Theory}

\subsection{Background}
In a FLRW universe fulfilled  with a pressureless component interacting with a vacuum-like term, the Friedmann and conservation equations assume the form
\begin{eqnarray} \label{Friedmann}
3H^2 = \rho_m + \Lambda,\\ \label{conservation}
\dot{\rho}_m + 3H\rho_m = \Gamma \rho_m = -\dot{\Lambda},
\end{eqnarray}
where $\Gamma$ is defined as the rate of matter creation (not necessarily constant). We will use a parametrisation for the vacuum term evolution given by
\begin{equation} \label{Lambda}
\Lambda = \sigma H^{-2\alpha},
\end{equation}
where $\alpha~(> -1)$ is the interaction parameter, and $\sigma = 3 (1 - \Omega_{m0}) H_0^{2(\alpha+1)}$. From (\ref{Friedmann}) and (\ref{conservation}) we can show that
\begin{equation} \label{Gamma}
\Gamma = -\alpha \sigma H^{-(2\alpha +1)}.
\end{equation}
By including a conserved radiation component, we obtain the Hubble function
\begin{equation}\label{eq:E}
E(z) = {H(z)}/{H_0} = \sqrt{\left[ (1-\Omega_{m0}) + \Omega_{m0} (1+z)^{3(1+\alpha)} \right]^{\frac{1}{(1+\alpha)}} + \Omega_{R0} (1+z)^4}.
\end{equation}
A negative $\alpha$ corresponds to creation of matter, while a positive one means that dark matter is annihilated. As one may check, for $\alpha = 0$  the standard $\Lambda$CDM model is recovered. Eq.~(\ref{eq:E}) is the Hubble function of a generalized Chaplygin gas (GCG) \cite{cg1,Dev:2002qa,Alcaniz:2002yt,cg2,cg3}, which behaves like cold matter at early times and as a cosmological constant in the asymptotic future. Actually, it is equivalent to a non-adiabatic Chaplygin  gas \cite{Borges:2013bya,non-adiabatic2,wands2,bb}, because the vacuum component does not cluster and, consequently, there is no pressure term in the perturbation equations. For this reason, the power spectrum does not suffer from oscillations and instabilities present in the adiabatic version. Note as well that conserved baryons are included in the term proportional to $(1+z)^3$ in the binomial expansion of the square brackets. A small spatial curvature can also be included by adding a term $\Omega_{k} (1+z)^2$ into the square root.

\subsection{Primordial perturbations}

The Boltzmann equations for conserved baryons and radiation are the same as in the standard model. For the dark sector, assuming that there is no momentum transfer in the dark matter rest frame, the Poisson and dark matter perturbation equations in the conformal Newtonian gauge are \cite{micol,Salzano:2021zxk}
\begin{equation}\label{thetad2}
\theta'_{dm}+\mathcal{H}\theta_{dm}-k^2\Phi=0,
\end{equation}
\begin{equation}\label{deltad2}
\delta'_{dm}-3\Phi'+\theta_{dm}=-\frac{aQ}{\rho_{dm}} \left[ \delta_{dm} - \frac{1}{k^2} \left( k^2 \Phi + \frac{Q'}{Q}\theta_{dm} \right) \right],
\end{equation}
\begin{equation}
-k^2 \Phi=\frac{a^2}{2} (\rho_{dm} \delta_{dm} + \rho_b \delta_b) - \left( \frac{a^3 Q}{2} - \frac{3a^2}{2}\mathcal{H}\rho_m \right)\frac{\theta_{dm}}{k^2},
\end{equation}
where $\mathcal{H} = aH$, $Q = \Gamma \rho_m = - \dot{\Lambda}$, a prime indicates derivative w.r.t. the conformal time, $\theta_{dm}$ is the dark matter velocity potential, and $\Phi$ is the gravitational potential.
In the sub-horizon limit $k \gg \mathcal{H}$, these equations assume the form
\begin{equation}\label{ol}
\theta'_{dm}+\mathcal{H}\theta_{dm}-k^2\Phi=0,
\end{equation}
\begin{equation}\label{ol1}
\delta'_{dm}+\theta_{dm}=-\frac{aQ}{\rho_{dm}} \delta_{dm},
\end{equation}
\begin{equation}\label{ol2}
-k^2 \Phi=\frac{a^2}{2} (\rho_{dm} \delta_{dm} + \rho_b \delta_b).
\end{equation}
For the vacuum term we have $\delta \Lambda = -aQ\theta_{dm}/k^2$ and $\delta Q = Q'\theta_{dm}/k^2$, that are negligible for sub-horizon modes. The vacuum term velocity remains undetermined.

\subsubsection{Primordial perturbations: non-flat case}

Let us show now that the inclusion of spatial curvature in the dynamical equations at late time is only relevant at the background level.
For this, consider the components of Einstein's equations in the longitudinal gauge,
\begin{equation}\label{omj}
\psi-\phi=a^2\pi,
\end{equation}
\begin{equation}\label{del}
3\mathcal{H}(\psi'+\mathcal{H}\phi)+k^2\psi-3\kappa\phi=-\frac{a^2}{2}\delta\rho,
\end{equation}
\begin{equation}\label{sees}
\psi''+2\mathcal{H}\psi'+\mathcal{H}\phi'+(2\mathcal{H'}+\mathcal{H}^2-\kappa)\phi=\frac{a^2}{2}(\delta p-\frac{2}{3}k^2\pi),
\end{equation}
\begin{equation}\label{kal}
\psi'+\mathcal{H}\phi=-\frac{a^2}{2}(\rho+p)v,
\end{equation}
where the spatial curvature can assume values $\kappa=0, +1$ ou $-1$. In the absence of anisotropic stress $\pi=0$, the gravitational potential and curvature perturbation are equal, $\phi=\psi$.
The perturbation in the fluid pressure is
\begin{equation}\label{1}
\delta p=c_{s}^2\delta\rho+(c_{s}^2-c_{a}^2)\rho'v,
 \end{equation} 
where $c_{s}^2$ and $c_{a}^2=p'/\rho'$ are, respectively, the entropic and adiabatic sound speed, and $v$ is the matter velocity potential.

With the help of the continuity equation, 
\begin{equation}\label{2}
\rho'=-3\mathcal{H}(\rho+p),
\end{equation}
we are able to rewrite the equation $(\ref{kal})$ as
\begin{equation}\label{3}
\frac{a^2}{2}\rho'v=3\mathcal{H}(\phi'+\mathcal{H}\phi).
\end{equation}
Substituting $(\ref{1})$ into $(\ref{sees})$ and eliminating $\delta\rho$ and $(a^2/2)\rho'v$ through equations $(\ref{del})$ and $(\ref{3})$, it is easy to find the second order differential equation for the gravitational potential,

\begin{equation}\label{tam}
\phi''+3\mathcal{H}(1+c_{a}^2)\phi'+[2\mathcal{H}'+(1+3c_{a}^2)\mathcal{H}^2+c_{s}^2(k^2-3\kappa)-\kappa]\phi=0.
\end{equation}
Combining $(\ref{kal})$ with $(\ref{del})$ and using the continuity equation $(\ref{2})$ we obtain the Poisson equation
\begin{equation}\label{gf}
-2(k^2-3\kappa)\phi=a^2\rho\delta^c,
\end{equation}
where $\delta\rho^{c}=\delta\rho+\rho'v$ is a comoving gauge invariant quantity.

For any interacting model with 
\begin{equation}
\rho_m'+3\mathcal H\rho_m=aQ,
\end{equation}
the adiabatic sound speed that appear in equation $(\ref{tam})$ is directly related to the energy transfer function as
\begin{equation}
c_a^2=-\frac{aQ}{3\mathcal H\rho_m}.
\end{equation}
As in the flat case, it is easy to show that the vacuum energy component does not cluster in sub-horizon scales, and therefore there is no entropic sound speed, i.e.
$c_s^2\propto \delta\rho_{\Lambda}^c=0$
in the perturbation equation $(\ref{tam})$. Then, using ($\ref{gf}$) into $(\ref{tam})$ we can find a second order differential equation for the evolution of the density contrast,
\begin{equation}\label{dif}
\delta_m^{c''}+\bigg(\mathcal H+\frac{aQ}{\rho_m}\bigg)\delta_m^{c'}+\bigg[\bigg(\frac{aQ}{\rho_m}\bigg)'+\frac{aQ}{\rho_m}\mathcal H+\mathcal H'-\mathcal H^2-\kappa\bigg]\delta_m^c=0.
\end{equation}
From the Friedmann and Raychaudhuri equations
\begin{equation}
\mathcal H^2+\kappa=\frac{a^2}{3}(\rho_m+\rho_{\Lambda}),
\end{equation}
\begin{equation}
\mathcal H'+\mathcal H^2+\kappa=\frac{a^2}{6}(\rho_m+4\rho_{\Lambda}),
\end{equation}
we derive the following relation,
\begin{equation}\label{sa}
\mathcal H'-\mathcal H^2-\kappa=-\frac{1}{2}a^2\rho_m.
\end{equation}
Note that the spatial curvature parameter $\kappa$ that appears in the last term of equation ($\ref{dif}$), due to the contribution of non-flat terms of Einstein's equations ($\ref{del}$) and ($\ref{sees}$), is eliminated in view of ($\ref{sa}$). Hence we can get an equation for $\delta_m^c$ in the final form
\begin{equation}
\delta_m^{c''}+\bigg(\mathcal H+\frac{aQ}{\rho_m}\bigg)\delta_m^{c'}+\bigg[\bigg(\frac{aQ}{\rho_m}\bigg)'+\frac{aQ}{\rho_m}\mathcal H-\frac{1}{2}a^2\rho_m\bigg]\delta_m^c=0.
\end{equation}
This same differential equation for total matter can be derived by combining equations ($\ref{ol}$)-($\ref{ol2}$), which assumes a universe with zero spatial curvature. Therefore, we see that the evolution of $\delta_m$ at sub-horizon scales is affected by spatial curvature only through its background solutions.

\section{Analysis and Results}
\label{Sec:Analysis}

For our analysis, we updated the CMB data set used in the previous work\footnote{In Ref.\cite{micol} 
we have used the second release of Planck data \cite{Aghanim:2015xee} ``TT+lowP" (2015), namely the high-$\ell$ Planck temperature data (in the range $30< \ell <2508$) from the 100-, 143-, and 217- GHz half-mission TT cross-spectra, and the low-P data by the joint TT, EE, BB and TE likelihood (in the range $2< \ell <29$).}, joining the Plik ``TT,TE,EE+lowE" CMB Planck (2018) likelihood (by combination of temperature power spectra and cross correlation TE and EE over the range $\ell \in [30, 2508]$, the low-$\ell$ temperature Commander likelihood, and the low-$\ell$ SimAll EE likelihood) \cite{Aghanim:2019ame}, its lensing reconstruction power spectrum\footnote{As shown by the Planck Collaboration, lensing data is needed both to resolve the tension of the CMB data with a flat universe prediction, and to reduce the tension of the CMB with data from Redshift Space Distorsion and Weak Lensing.
Let us stress that the Planck ``TT,TE,EE+lowE + lensing" data set combination is considered by the Planck Collaboration as the most robust of the current release \cite{Aghanim:2018eyx}.} ~\cite{Aghanim:2019ame,Aghanim:2018oex} and SNe data from the Joint Light-curve sample~\cite{Betoule:2014frx}. The latter is constructed from Supernova Legacy Survey (SNLS) and Sloan Digital Sky Survey (SDSS), consisting of 740 data points covering the redshift range $0.01< z <1.3$. This sample allows for light-curve recalibration with the model under consideration, which is an important issue when testing alternative cosmologies~\cite{Taddei:2016iku,micol}. Although SNe data have lower statistical power with respect to Planck, they are useful for fixing the background cosmology at low redshifts in models involving dark energy evolution and modified gravity.
We include a prior on $\Omega_{b0}h^2$ in order to take into account the observations of  D/H abundance \cite{Cooke:2017cwo}, and we call such a data set Planck(2018)+lensing+JLA+$\Omega_{b0}h^2$ prior as ``base data set".
 Also, we consider the Hubble constant of SH$0$ES collaboration, $H_0$ = $74.03 \pm 1.42$ km/s/Mpc ~\cite{Riess:2019cxk}, that is in tension at 4.4$\sigma$ with CMB estimations within the minimal cosmological model, to discuss the changes in the parameters constraining due to the assumption of this prior.
\begin{table}[]
\centering
\caption{{
$68\%$ confidence limits for the model parameters. We call as ``base data set" the set of  Planck(2018)+lensing+JLA+$\Omega_{b0}h^2$ prior. 
The $\Delta \chi^2_{best} = \Delta \chi^2_{\Lambda CDM} - \Delta \chi^2_{model}$ refers to the best fit of the model (negative value means a better $\chi^2$ of the reference model, $\Lambda$CDM).}
\label{tab:flat}}
\begin{tabular}{|c|c|c|c|c|}
\hline
\multicolumn{1}{|c|}{Parameter}&
\multicolumn{2}{c}{$\Lambda$CDM}&
\multicolumn{2}{|c|}{$\Lambda$(t)CDM}\\
\hline
{ }&
{base dataset }&
{base dataset + SH$0$ES}&
{base dataset} & 
{base dataset + SH$0$ES}\\ 
\hline
$100\,\Omega_b h^2$ 	
& $2.223 \pm 0.012$ %
& $2.235 \pm 0.012$ %
& $2.221 \pm 0.012$ %
& $2.221 \pm 0.013$ %
\\
$\Omega_{cdm} h^2$	
& $0.1206 \pm 0.0012$ %
& $0.1190 \pm 0.0011$ %
& $0.1156 \pm 0.0099$ %
& $0.0890 \pm 0.0093$ 
\\
$\alpha$	
& $-$ %
& $-$ %
& $0.03 \pm 0.06$ %
& $0.20 \pm 0.07$ 
\\
$H_0$
& $ 67.04 \pm 0.50 $ 
& $ 67.78 \pm 0.48 $ 
& $ 67.50 \pm 1.22 $ 
& $ 70.73 \pm 1.02 $ %
\\
\hline
\hline
$\Delta \chi^2_{\rm best}$         
& $-$ 
& $-$ 
& $0.8 $ 
& $ 9.3$ 
\\
\hline
\end{tabular}
\end{table} 

\begin{table}[]
\centering
\caption{{
$68\%$ confidence limits for the model parameters, using the ``base dataset".
The $\Delta \chi^2_{best} = \Delta \chi^2_{\Lambda CDM} - \Delta \chi^2_{model}$ refers to the best fit of the model (negative value means a better $\chi^2$ of the reference model, $\Lambda$CDM).}
\label{tab:curve}}
\begin{tabular}{|c|c|c|}
\hline
{Parameter}&
{$\Lambda$CDM + $\Omega_k$}& 
{$\Lambda$(t)CDM + $\Omega_k$}\\ 
\hline
$100\,\Omega_b h^2$ 	
& $2.220 \pm 0.012$ %
& $2.225 \pm 0.013$ %
\\
$\Omega_{cdm} h^2$	
& $0.1208 \pm 0.0011$ %
& $0.1123 \pm 0.0177$ %
\\
$\Omega_{k}$	
& $0.0008 \pm 0.0008$ %
& $-0.004 \pm 0.007$ %
\\
$\alpha$	
& $-$ %
& $0.06 \pm 0.11$ 
\\
$H_0$
& $ 67.29 \pm 0.60 $ 
& $ 66.38 \pm 1.94 $ 

\\
\hline
\hline
$\Delta \chi^2_{\rm best}$         
& $ - $ 
& $ -0.4 $ 
\\
\hline
\end{tabular}
\end{table} 

 We modify the numerical Cosmic Linear Anisotropy Solving System (CLASS) code~\cite{Blas:2011rf} according with the theory  discussed in the previous sections, and we use the Monte Python~\cite{Audren:2012wb} code to perform Monte Carlo Markov Chains (MCMC) analyses. We build our theory by letting free the usual cosmological parameters, namely, the physical baryon density, $\omega_b=\Omega_{b}h^2$, the physical cold dark matter density, $\omega_{cdm}=\Omega_{cdm}h^2$, the optical depth, $\tau_{reio}$, the primordial scalar amplitude, $\mathcal A_s$, the primordial spectral index, $n_s$, the Hubble constant $H_0$, in addition to the interaction parameter, $\alpha$. 
%
%
\begin{figure}[t]
\centerline{\includegraphics[scale=0.4]{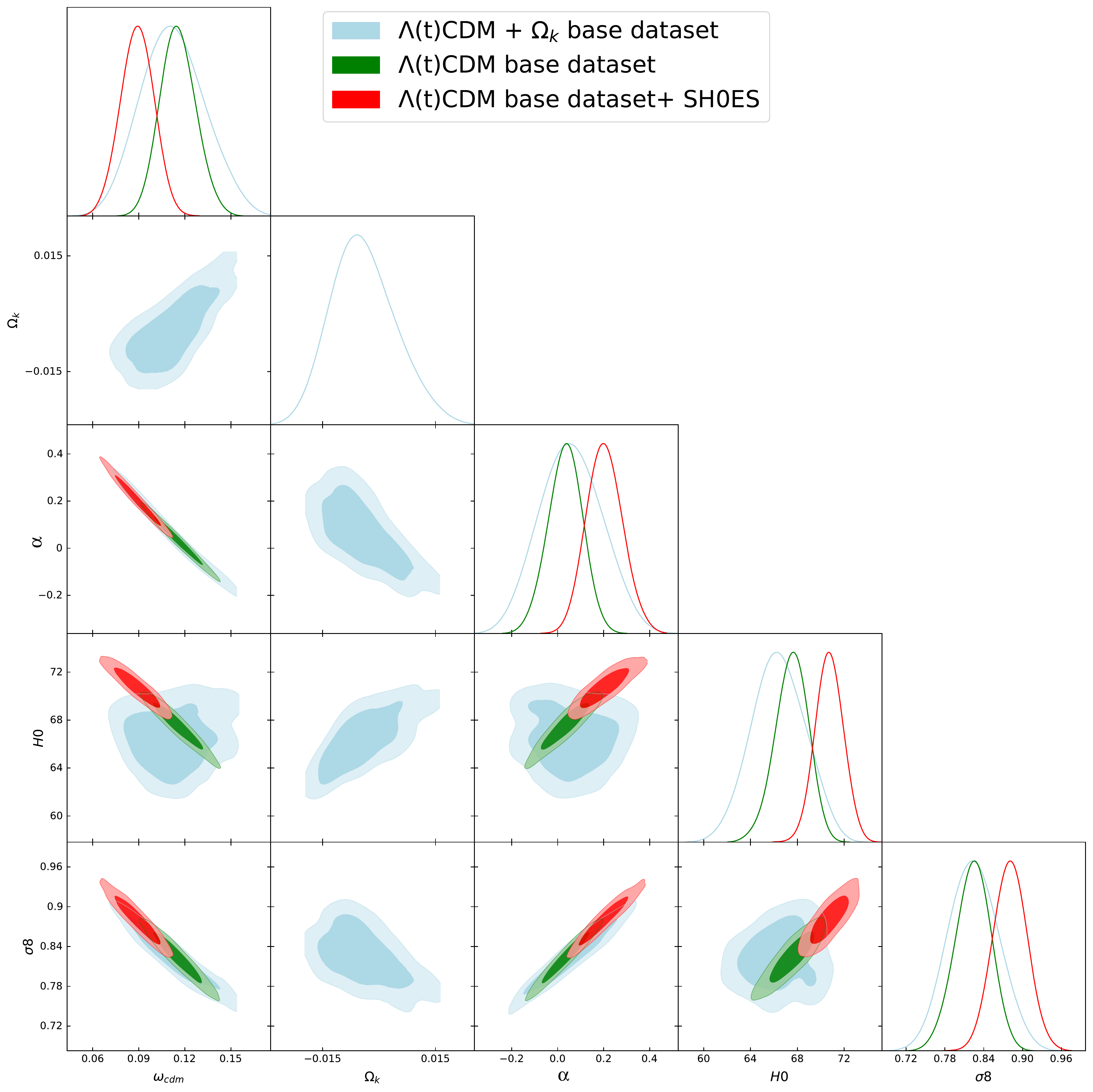} \hspace{.4in}}
\caption{Comparison between flat and curved $\Lambda$(t)CDM models, using as ``base dataset" the Planck 2018 likelihood combined with SNe JLA sample and $\Omega_{b0}h^2$ prior.}
\label{fig:triangle_LtCDM}
\end{figure}

We present our results in Table \ref{tab:flat} (flat universe) and Table \ref{tab:curve} (non-zero curvature). 
Within the standard model framework, we first note that the base data set fully supports the flat hypothesis. 
At the same time, when a Gaussian prior on $H_0$ centered in the SH$0$ES value is considered, we note a shift in the constrained value of the cold dark matter density, although the two values remain within $1\sigma$ agreement. 
This behavior is not observed when we analyse the $\Lambda$(t)CDM model, where the $H_0$ prior has a significant influence both on the peak shift of $\omega_{cdm}$ to lower values (removing such a prior it is fully compatible with that predicted by the standard model) and on the $\alpha$ constraint. In this latter, using the SH$0$ES prior the standard model is discarded at $3\sigma$. These results are shown in Fig. \ref{fig:triangle_LtCDM}, where we show the $\Lambda$(t)CDM model with (red line) and without (green line) the SH$0$ES prior, and without the assumption of flat universe (light blue line).

The analysis considering a curved space in the context of the $\Lambda$(t)CDM model shows an anti-correlated behaviour between the curvature density and the $\alpha$ parameter. A flat and non-interacting universe is still compatible at $1\sigma$ even if the data show preference for slight negative curvature values and slight positive interaction parameter. 
Noteworthy, it seems that a spatially curved universe relaxes the degeneracy between $\alpha$ and $H_0$, and also between $H_0$ and $\sigma_8$.

\begin{figure}[t]
\centerline{\includegraphics[scale=0.3]{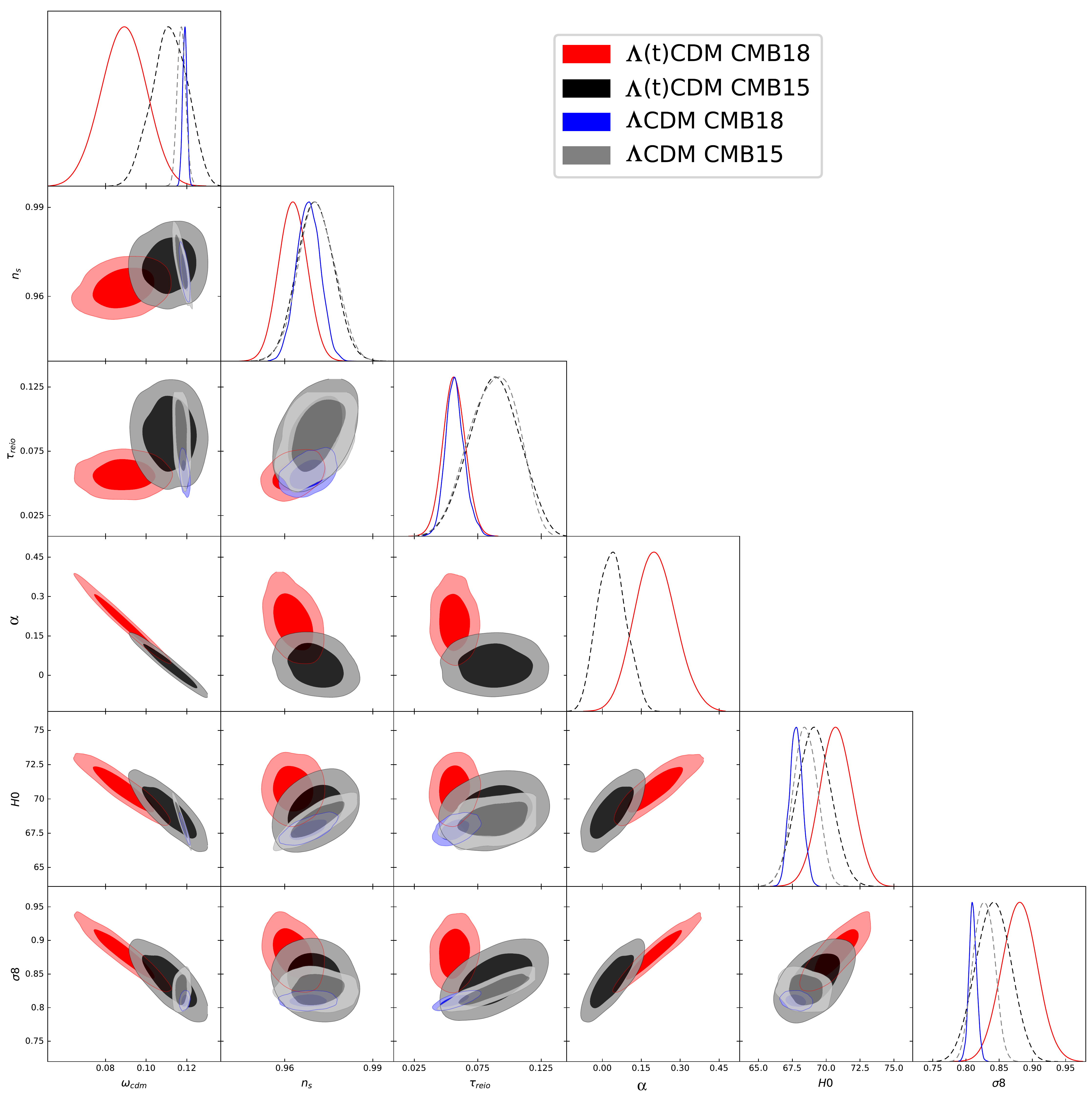} \hspace{.4in}}
\caption{Comparison between $\Lambda$CDM and $\Lambda$(t)CDM models using Planck likelihood 2015 and Planck likelihood 2018 combined with SNe JLA sample, as well as $\Omega_{b0}h^2$ and SH0ES priors.}
\label{fig:triangle_LtCDMvsLCDM}
\end{figure}

Finally, in order to comment our results in the light of the previous ones \cite{micol}, let us now compare the results obtained for both $\Lambda$(t)CDM and standard models using both 2015 and 2018 Planck likelihoods, combined with JLA + $\Omega_{b0}h^2$ prior + SH0ES prior. We remind that the analysis using Planck 2015 (dashed black curve) was presented in \cite{micol}, where we used the most robust combination at the time, namely ``TT + lowP" \cite {Aghanim:2015xee}, while for the new analysis of this work we have chosen the combination currently considered more reliable, that is ``TT,TE,EE + lowE" \cite{Aghanim:2019ame}. The main difference between the two Planck likelihoods essentially lies in: (i) the use at high-$\ell$ of polarization modes, and (ii) a different treatment of EE polarization at low multipoles \cite{Aghanim:2019ame}, that implies a stronger constraint on the optical depth parameter (for a detailed discussion we refer the reader to Ref. \cite{Aghanim:2018eyx}). Due to the correlation between cosmological parameters, this also determines a preference, in the standard cosmological model context, for lower values of $A_s$ and the late-time fluctuation amplitude parameter, $\sigma_8$\footnote{The $\sigma_8$ parameter roughly corresponds to the primordial scalar amplitude, $A_s$, converted into present fluctuations amplitude.}, than those estimated by Planck (2015) likelihood.
This can be seen in Fig. \ref{fig:triangle_LtCDMvsLCDM}, comparing the standard model constrained with Planck data from the 2015 release (dashed gray curve) and the 2018 release (solid blue curve).
At the same time, we also note a significant different constraint on the cold dark matter density, comparing the $\Lambda$(t)CDM (red solid curve) and $\Lambda$CDM model using 2018 CMB data. On the other hand, the Hubble parameter is compatible with that predicted by $\Lambda$CDM at only $2.6\sigma$, relaxing the tension with the value measured by SH$0$ES to 1.9$\sigma$. We also emphasize that $\alpha=0$ (the value of the interaction parameter required to recover the standard model) is excluded by the analysis using the 2018 CMB data, with a clear preference for a positive $\alpha$, that is, an energy flux from dark matter to dark energy.

As noted in the previous work \cite{micol}, there exist a positive correlation between the values of $\alpha$ and $H_0$, which implies (due to the additional degree of freedom) that the $\Lambda$(t)CDM model can predict values of $H_0$ in better agreement with  the local measurements than the standard $\Lambda$CDM model. Our present analysis not only confirms this correlation but also shows that it is even stronger when the Planck (2018) data are used.  On the other hand, given the anti-correlation between these two latter  parameters and $\omega_{cdm}$,  a significant shift of the cold dark matter density to smaller values are obtained for the iDE model when compared with the standard model prediction. It is worth noticing that these results are obtained using a prior on $H_0$ given by the SH$0$ES collaboration. By removing this prior from the analysis the constraints on these parameters are compatible with those of the $\Lambda$CDM model, as shown in Fig. \ref{fig:triangle_LtCDM}.

\section{Final Remarks}
\label{Sec:Conclusions}
In this paper we have not only updated the previous results of \cite{micol} with the Planck 2018 polarization data, but also explored the influence of non-zero spatial curvature, and the weight of the priors choice in the analysis, particularly the use of the SH$0$ES prior on the local value of the Hubble parameter, when constraining the cosmological parameters with Planck data. Taking such a $H_0$ prior in combination with CMB data has in fact opened up numerous debates on whether it is statistically valid or not to perform  analyses of models by combining data in tension \cite{Verde:2013wza, Handley:2019wlz, Battye:2014qga, Seehars:2014ora, Raveri:2018wln, Efstathiou:2020wem, Gonzalez:2021}. 

The choice to use this prior must therefore be seriously considered and the results obtained carefully analysed. We have shown that, without such a prior, the current CMB data are not capable of discerning an interaction in the dark sector, even in combination with SNe Ia data. On the other hand, when the SH$0$ES prior is taken into account, the  best-fit of the $\alpha$ parameter is clearly positive, whereas the standard model ($\alpha=0$) is excluded with $\approx 3\sigma$ confidence level. This results from the fact that the interaction and the Hubble parameters are directly correlated, as shown by the analysis with Planck data only. Therefore, a prior that prefers a higher value for the latter will also naturally lead to a higher value to the former.

We have also considered the role of spatial curvature in the data analysis.  We note that the best-fits of the cosmological parameters are not substantially altered, and no robust sign of the presence of curvature can be concluded. In this respect, the curvature has a weaker influence on the analysis as compared to the number of relativistic species, that leads to a negative interaction parameter if left free, as shown in \cite{micol}. Our general conclusion is that the signature of interaction, if exists, is too weak to be found with the present data set. On the other hand, the present analysis also suggests that the $H_0$ tension observed in the context of the $\Lambda$CDM model seems to be more fundamental, refusing solutions based exclusively on generalizations of the dark sector.

\section*{Acknowledgements}

We are thankful to Joel Carvalho for an insightful discussion on the proper use of the SH$0$ES prior. MB thanks support of the Istituto Nazionale di Fisica Nucleare (INFN), sezione di Napoli, iniziative specifiche QGSKY. SC is supported by CNPq (Brazil) with grant 307467/2017-1. JSA acknowledges support from CNPq (grants no.~310790/2014-0 and 400471/2014-0) and FAPERJ (grant no.~204282). The authors thank the use of CLASS and Monte Python codes. We also acknowledge the use of the High Performance Data Center (DCON) at the Observat\'orio Nacional for providing the computational facilities to run our analysis.

\end{document}